\documentclass[aps,prb,showpacs,preprint,amsfonts,amssymb,amsmath]{revtex4}
\usepackage{amsmath,amssymb}
\usepackage{dcolumn}
\usepackage{bm}

\begin{document}

\title{Double quantum dot as detector of spin bias}
\author{Qing-feng Sun$^{1}$, Yanxia Xing$^{1}$, and Shun-Qing Shen$^{2}$}
\affiliation{$^1$Beijing National Lab for Condensed Matter Physics
and Institute of Physics, Chinese Academy of Sciences, Beijing
100080, China
\\
$^2$Department of Physics, and Center of Theoretical and
Computational Physics, The University of Hong Kong, Hong Kong, China
}
\date{\today}

\begin{abstract}
It was proposed that a double quantum dot can be used to be a
detector of spin bias. Electron transport through a double quantum
dot is investigated theoretically when a pure spin bias is applied
on two conducting leads contacted to the quantum dot. It is found
that the spin polarization in the left and right dots may be
induced spontaneously while the intra-dot levels are located
within the spin bias window and breaks the left-right symmetry of
the two quantum dots. As a result, a large current emerges. For an
open external circuit an charge bias instead of a charge current
will be induced in equilibrium, which is believed to be measurable
according to the current nanotechnology. This method may provide a
practical and whole electrical approach to detect the spin bias
(or the spin current) by measuring the charge bias or current in a
double quantum dot.
\end{abstract}

\maketitle

\section{Introduction}

Discovery and application of giant magnetoresistance (GMR) in metallic thin
films marks the beginning of a new era of spintronics.\cite{ref1,ref2} Since
then, people begin to exploit electron spin to replace the role of electron
charge in electronic devices. As a counterpart of charge current, spin
current, in which spin-up and spin-down electrons move coherently in
opposite directions, has been attracted extensive interests.\cite{Awschalom}
Various methods were proposed to generate spin current,\cite{ref3} and to
explore the characteristics of the spin transport. Over last few years,
search of spin current has made a great of progresses. It has been generated
and detected successfully by various means, such as the optical injection,%
\cite{ref4, cui} the magnetic tunnelling injection,\cite{ref5,Kimura} or the
spin Hall effect.\cite{ref6,ref7} All these experiments focus on the optical
measurement of spin accumulation near the boundaries of sample or electric
measurement of the scattering effect induced by the spin current via
spin-orbital coupling. There are also some proposals to measure spin current
or spin polarized current,\cite{ref8,addref1,ref9,add1} e.g. to measure the
spin torque while a spin current flowing through a ferromagnetic-nonmagnetic
interface,\cite{ref8} or to detect the induced electric field by the spin
current.\cite{addref1,ref9} In all these methods, it always involves the
optical, magnetic materials or impurities, magnetic field, or spin-orbit
interaction. Up to now, it is still a challenge to detect the spin current
efficiently, which has become a bottleneck of the development of the
spintronics.

When a spin current flows through a device, there always exists a spin bias
between the two terminals of the device.\cite{addnote1} A spin bias means
that the chemical potentials of the two terminals are spin-dependent (see
Fig.1). The spin bias is regarded as the driving force behind the spin
current. When the circuit is open, the spin current has to be zero.
Consequently the spin bias usually induce spin accumulation in equilibrium.
When the circuit is connected, a spin current circulates. The relation
between spin bias and spin current is very similar with the relation between
the charge bias and charge current. On the charge transport, people often
detect the charge bias to replace the measurement of the charge current.
Correspondingly we can also measure the spin bias instead of the spin
current. In this paper, we propose an effective method to detect the spin
bias.

The present proposal is a whole electric measurement of spin bias by means
of a double quantum dot (DQD). It does not involve any optical or magnetic
means, and even the spin-orbit interaction. The spin bias can be detected by
measuring the (charge) bias. The DQD can be regarded as an artificial
molecule, and the electron numbers in DQD can be controlled very well. In
last two decades, the electron transport through the DQD device has been
extensively investigated.\cite{ref10,ref11} DQD has also been proposed as a
qubit,\cite{ref12} a device to detect various tunnelling rates and spin flip
rate,\cite{ref11,ref13} and so on. Here we propose that a DQD can be applied
to measure the spin bias or spin current.

Let us first describe the working mechanism of DQD as a detector of spin
bias. Consider a DQD coupled into two conducting leads. Suppose a spin bias
be applied between the left and right leads. Our task is to measure this
spin bias \textsl{experimentally}. The spin bias is defined as the
spin-dependent chemical potentials of the two leads with $\mu _{L\uparrow
}=-\mu _{L\downarrow }=-\mu _{R\uparrow }=\mu _{R\downarrow }=V$ (see Fig.1).%
\cite{ref14} Assume that the left-dot level $\epsilon _{L}$ is set at zero
and the right-dot level $\epsilon _{R}$ is at $-U$, where $U$ is the
intra-dot electron-electron (e-e) Coulomb interaction. This particular level
position is chosen to demonstrate the physics in our proposal, and is not
necessary at all in a general case. The left dot has a spin-up electron
because of $\mu _{L\uparrow }>\epsilon _{L}>\mu _{L\downarrow }$, while the
right dot, because of $\mu _{R\downarrow }>\epsilon _{R}+U>\mu _{R\uparrow
}>\epsilon _{R}$, is occupied by a spin-down electron, and its spin-up level
is consequently pushed away to the higher energy $\epsilon _{R}+U$ and is
empty (see Fig.1). The spin-up electron can then tunnel from the left lead
via the two dots to the right lead (see Fig.1a). Oppositely the spin-down
electron can hardly flow from the right lead to the left lead because of the
Pauli exclusion principle and the occupancy of the spin-down level in the
right dot (see Fig.1b). This breaks the symmetry of the motion of spin-up
and spin-down electrons in a pure spin bias. As a result, a (charge) current
circulates. This induced current can be measured experimentally, and
consequently be applied to measure the spin bias.

The paper is organized as follows. In Section II, the model for the
DQD and the general formalism for nonequilibrium Keldysh Green's
function method are presented. The spin-bias-induced charge current
$J$ and the electron occupation numbers in the DQD are calculated.
In Section III, we take the numerical investigation. The
spin-dependent charge stability diagram in terms of the spin bias is
obtained. In Section IV, the induced charge bias in an open circuit
is numerically studied. Finally, a brief summary is presented in
Section V.

\section{model and formulation}

In this section, we present the model Hamiltonian of this DQD and the
general formalism of Keldysh Green's function technique for electron
transport through the DQD. The DQD device is modelled by the following
Hamiltonian,
\begin{eqnarray}
H &=&\sum\limits_{\alpha ,k,\sigma }\epsilon _{\alpha k}a_{\alpha k\sigma
}^{\dagger }a_{\alpha k\sigma }+\sum\limits_{\alpha ,\sigma }\epsilon
_{\alpha }d_{\alpha \sigma }^{\dagger }d_{\alpha \sigma }  \notag \\
&&+\sum\limits_{\alpha }U_{in}d_{\alpha \uparrow }^{\dagger }d_{\alpha
\uparrow }d_{\alpha \downarrow }^{\dagger }d_{\alpha \downarrow
}+\sum\limits_{\sigma ,\sigma ^{\prime }}U_{ex}d_{L\sigma }^{\dagger
}d_{L\sigma }d_{R\sigma ^{\prime }}^{\dagger }d_{R\sigma ^{\prime }}  \notag
\\
&&+\sum\limits_{\alpha ,k,\sigma }t_{\alpha }a_{\alpha k\sigma }^{\dagger
}d_{\alpha \sigma }+\sum_{\sigma }t_{c}d_{L\sigma }^{\dagger }d_{R\sigma
}+H.c.
\end{eqnarray}%
where $a_{\alpha k\sigma }^{\dagger }$ ($a_{\alpha k\sigma }$) and $%
d_{\alpha \sigma }^{\dagger }$ ($d_{\alpha \sigma }$) are the creation
(annihilation) operators of electron with spin $\sigma $($=\uparrow
,\downarrow )$ in the lead $\alpha $($=L,R)$ and the dot $\alpha $ ,
respectively. Each dot has a single energy level $\epsilon _{\alpha }$ and
an intra-dot e-e interaction $U_{in}$. In addition, the inter-dot e-e
interaction $U_{ex}$ is also included. We emphasize that the system does not
break the spin SU(2) symmetry, and the hopping coefficients $t_{\alpha }$
and $t_{c}$ are spin-independent.

Following the transport theory of Keldysh Green's function,\cite{ref15} the
electron current $J_{\alpha \sigma }$ with the spin $\sigma $ from the lead $%
\alpha $ flowing into the dot $\alpha $ and the occupation number of
electron $n_{\alpha \sigma }$ at the level $\alpha ,\sigma $ can be
expressed as,
\begin{eqnarray}
J_{\alpha \sigma } &=&-Im\int \frac{d\epsilon }{2\pi }\Gamma _{\alpha }\left[
2f_{\alpha \sigma }G_{\alpha \alpha \sigma }^{r}(\epsilon )+G_{\alpha \alpha
\sigma }^{<}(\epsilon )\right] \\
n_{\alpha \sigma } &=&\langle d_{\alpha \sigma }^{\dagger }d_{\alpha \sigma
}\rangle =-i\int \frac{d\epsilon }{2\pi }G_{\alpha \alpha \sigma
}^{<}(\epsilon )
\end{eqnarray}%
where $\Gamma _{\alpha }=2\pi \sum_{k}|t_{\alpha }|^{2}\delta (\epsilon
-\epsilon _{\alpha k})$. $f_{\alpha \sigma }(\epsilon )=1/\{exp[(\epsilon
-\mu _{\alpha \sigma })/k_{B}T]+1\}$ is the Fermi-Dirac distribution of
electrons in the leads. Because of the spin bias in the two leads, the
chemical potentials for spin-up and spin-down electrons are not equal. $%
G_{\alpha \alpha \sigma }^{r}(\epsilon )$ and $G_{\alpha \alpha \sigma
}^{<}(\epsilon )$ in Eqs. (2) and (3) are the standard retarded and the
Keldysh Green's functions of the QDs, they are the Fourier transformation of
$G_{\alpha \alpha \sigma }^{r,<}(t)$, where
\begin{eqnarray*}
G_{\alpha \alpha ^{\prime }\sigma }^{r}(t) &\equiv &-i\theta (t)\langle
\{d_{\alpha \sigma }(t),d_{\alpha ^{\prime }\sigma }^{\dagger }(0)\}\rangle ,
\\
G_{\alpha \alpha ^{\prime }\sigma }^{<}(t) &\equiv &i\langle d_{\alpha
\sigma }^{\dagger }(0)d_{\alpha ^{\prime }\sigma }(t)\rangle .
\end{eqnarray*}

We first solve the Green's functions $\mathbf{g}_{\sigma }^{r}(\epsilon )$
of the isolated DQDs system (i.e. $t_{\alpha }=t_{c}=0$). Consider that the
spin bias $V$ is less than the intra-dot e-e interaction $U_{in}$ and the
two-electron co-tunneling events can be ignored. $\mathbf{g}_{\sigma
}^{r}(\epsilon )$ are obtained from the equation of motion technique:\cite%
{note21}
\begin{eqnarray}
g_{\alpha \alpha \sigma }^{r}(\epsilon ) &=&\frac{(1-n_{\alpha \bar{\sigma}%
})(1-\{n_{\bar{\alpha}}\})}{A}+\frac{(1-n_{\alpha \bar{\sigma}})\{n_{\bar{%
\alpha}}\}}{A-U_{ex}}  \notag \\
&&+\frac{n_{\alpha \bar{\sigma}}(1-\{n_{\bar{\alpha}}\})}{A-U_{in}}+\frac{%
n_{\alpha \bar{\sigma}}\{n_{\bar{\alpha}}\}}{A-U_{in}-U_{ex}},
\end{eqnarray}%
and $g_{LR\sigma }^{r}=g_{RL\sigma }^{r}=0$, where $\bar{\alpha}=R$ for $%
\alpha =L$ and $\bar{\alpha}=L$ for $\alpha =R$, $\bar{\sigma}=\downarrow $
for $\sigma =\uparrow $ and $\bar{\sigma}=\uparrow $ for $\sigma =\downarrow
$, $A\equiv \epsilon -\epsilon _{\alpha }-[n_{\bar{\alpha}}]U_{ex}+i0^{+}$, $%
\{n_{\bar{\alpha}}\}\equiv n_{\bar{\alpha}}-[n_{\bar{\alpha}}]$, and $%
[n_{\alpha }]$ is the integer part of $n_{\alpha }$. $n_{\alpha }=n_{\alpha
\uparrow }+n_{\alpha \downarrow }$ is the total occupation number of
electron in the dot $\alpha $. After solving $\mathbf{g}_{\sigma
}^{r}(\epsilon )$ of the isolated DQDs, $G_{\alpha \alpha \sigma
}^{r}(\epsilon )$ and $G_{\alpha \alpha \sigma }^{<}(\epsilon )$ for the
whole system can be obtained from Dyson and Keldysh equations:\cite{addnote2}
\begin{eqnarray}
\mathbf{G}_{\sigma }^{r}(\epsilon )\equiv \left(
\begin{array}{ll}
G_{LL\sigma }^{r} & G_{LR\sigma }^{r} \\
G_{RL\sigma }^{r} & G_{RR\sigma }^{r}%
\end{array}%
\right) &=&\mathbf{g}_{\sigma }^{r}(\epsilon )+\mathbf{g}_{\sigma
}^{r}(\epsilon )\mathbf{\Sigma }_{\sigma }^{r}\mathbf{G}_{\sigma
}^{r}(\epsilon ), \\
\mathbf{G}_{\sigma }^{<}(\epsilon )\equiv \left(
\begin{array}{ll}
G_{LL\sigma }^{<} & G_{LR\sigma }^{<} \\
G_{RL\sigma }^{<} & G_{RR\sigma }^{<}%
\end{array}%
\right) &=&\mathbf{G}_{\sigma }^{r}(\epsilon )\mathbf{\Sigma }_{\sigma
}^{<}(\epsilon )\mathbf{G}_{\sigma }^{a}(\epsilon ).
\end{eqnarray}%
Here the bold face letters ($\mathbf{G}$, $\mathbf{g}$, and $\mathbf{\Sigma }
$) represent the $2\times 2$ matrix, and the self-energies $\mathbf{\Sigma }%
_{\sigma }^{r,<}(\epsilon )$ are:
\begin{eqnarray}
\mathbf{\Sigma }_{\sigma }^{r}(\epsilon ) &=&\left(
\begin{array}{cc}
-i\Gamma _{L}/2 & t_{c} \\
t_{c} & -i\Gamma _{R}/2%
\end{array}%
\right) , \\
\mathbf{\Sigma }_{\sigma }^{<}(\epsilon ) &=&\left(
\begin{array}{cc}
i\Gamma _{L}f_{L\sigma }(\epsilon ) & 0 \\
0 & i\Gamma _{R}f_{R\sigma }(\epsilon )%
\end{array}%
\right) .
\end{eqnarray}%
Eqs. (3, 4, 5, and 6) can be solved self-consistently. The (charge) current
through the DQD is given by
\begin{equation*}
J=e(J_{L\uparrow }+J_{L\downarrow })=-e(J_{R\uparrow }+J_{R\downarrow }).
\end{equation*}

Finally it is worth pointing out that the present problem can be solved by
other means, for example,\ the rate equation method.\cite{addnote4}

\section{Spin-dependent Charge Stability Diagram and Charge Current}

Before presenting numerical results, we emphasize that the spin bias we
apply to the DQDs device is a pure symmetric one without a (charge) bias,
i.e. $\mu _{L\uparrow }+\mu _{L\downarrow }=\mu _{R\uparrow }+\mu
_{R\downarrow }=0$.\cite{ref14} So if the spontaneously spin-polarized
occupations are not induced in the DQD, the charge current $J$ must be zero
because of the symmetric behaviors for the motion of spin-up electron and
the spin-down electron. For example, in the case of a single quantum dot
instead of DQDs applied by the pure spin bias, there is no spin polarization
in the dot and the current is always zero as the spin up-down symmetry is
retained. So, in the following, we first investigate the stability diagram
of spin polarization and the spin-dependent charge density in the DQD.

Fig.2a and b present the spin polarizations $\Delta n_{\alpha }$ ($\Delta
n_{\alpha }\equiv n_{\alpha \uparrow }-n_{\alpha \downarrow }$) of the left
and right dots versus the levels $\epsilon _{L}$ and $\epsilon _{R}$, and
Fig. 2c presents the occupation number of electron $n_{L}+n_{R}/2$.\cite%
{note} It is found that these quantities are determined by the relative
energy levels of $\epsilon _{L}$ and $\epsilon _{R}.$ The spin polarization $%
\Delta n_{\alpha }$ is indeed non-zero and even quite large (i.e. near $\pm
1 $) in some specific regions. Let us analyze the spin-dependent charge
stability diagram (see Fig.3a), which gives spin-dependent occupation
numbers of electron as a function of $\epsilon _{L}$ and $\epsilon _{R}$. If
without the spin bias ($V=0$), there are four domains $(0,1)$, $(1,1)$, $%
(0,2)$, and $(1,2)$ in the stability diagram (see the thin dashed curves in
Fig.3a), with $(n,m)$ representing $n$ and $m$ electrons in the left and
right dot. This type of charge stability diagram has been observed
experimentally,\cite{ref10,ref11} and is well established. While the spin
bias $V$ is turned on and the level $\epsilon _{L}$ or $\epsilon _{R}$
locates between $-V$ and $+V$, in addition of the four old spin-unpolarized
domains $(n,m)$ with a shift $V$ of their boundaries shift, there appears
four new spin-polarized domains, which are denoted by $(\uparrow ,1)$, $%
(0,\downarrow )$, $(1,\downarrow )$, and $(\uparrow ,2)$. The notation $%
(\uparrow ,1)$, for example, represents an electron of spin-up in the left
dot and a spin-unpolarized electron in the right dot.

This spin-dependent charge stability diagram in Fig.3a can be obtained by
calculating the electrochemical potentials of the DQD or by analyzing the
level's position relative to the spin-dependent chemical potentials $\mu
_{\alpha \sigma }$. Consider the isolated DQDs device with $\Gamma
_{L}=\Gamma _{R}=t_{c}=0$. (i) The domain (0,1): when the equivalent level $%
\tilde{\epsilon}_{L}$ ($\tilde{\epsilon}_{L}\equiv \epsilon _{L}+U_{ex}$) of
the left dot is higher than $\mu _{L\uparrow }$ and $\mu _{L\downarrow },$
and the right-dot's level $\epsilon _{R}$ satisfies $\epsilon _{R}<\epsilon
_{L},\mu _{R\uparrow },\mu _{R\downarrow }<\epsilon _{R}+U_{in}$ (see
Fig.3b), the right dot is occupied by a spin-unpolarized electron and the
left dot is empty. (ii) The domain $(\uparrow ,1)$: while $\mu _{L\downarrow
}<\tilde{\epsilon}_{L}<\mu _{L\uparrow }$ and $\epsilon _{R}<\mu _{R\uparrow
},\mu _{R\downarrow }<\epsilon _{R}+U_{in}$ (see Fig.3c), a spin-up electron
occupies the left dot and a spin-unpolarized electron is in the right dot.
(iii) The domain $(0,\downarrow )$: if $\tilde{\epsilon}_{L}>\mu _{L\uparrow
},\mu _{L\downarrow }$ and $\mu _{R\uparrow }<\epsilon _{R}+U_{in}<\mu
_{R\downarrow }$ (see Fig.3d), the left dot is empty. For the right dot, a
spin-down electron occupies the level $\epsilon _{R}$ because of $\epsilon
_{R},\epsilon _{R}+U_{in}<\mu _{R\downarrow }$, then the spin-up level of
the right dot is pushed to $\epsilon _{R}+U_{in}$ which is over $\mu
_{R\uparrow }$, and so it is empty. Similarly, the other five domains can
also be obtained. In the case of the finite coupling case $\Gamma
_{L},\Gamma _{R},t_{c}\not=0$, the spin-polarized domains slightly extend to
the spin-unpolarized domains as illustrated in the thin dotted lines in
Fig.3a. Numerical results for the spin polarizations $\Delta n_{\alpha }$
(Fig.2a and b) and the occupation numbers of electrons $n_{L}+n_{R}/2$
(Fig.2c) are in a good agreement with the charge stability diagram in
Fig.3a. The eight domains, including four spin-unpolarized and four
spin-polarized domains, are clearly visible.

In an alternative way, the stability diagram of Fig.3a can also be deduced
from the total energy of the DQD system and the electrochemical potentials.
When the isolated DQD is in the states of $\vec{N}=(N_{L\uparrow
},N_{L\downarrow },N_{R\uparrow },N_{R\downarrow }),$ where $N_{\alpha
\sigma }=0$ or $1$ is the index of the electron occupation number in the
intra-dot level $\alpha \sigma $, its total energy $E_{T}$ is
\begin{eqnarray}
E_{T}(\vec{N}) &=&N_{L}\epsilon _{L}+N_{R}\epsilon _{R}+N_{L}N_{R}U_{ex}
\notag \\
&&+(N_{L\uparrow }N_{L\downarrow }+N_{R\uparrow }N_{R\downarrow })U_{in},
\end{eqnarray}%
with $N_{\alpha }=N_{\alpha \uparrow }+N_{\alpha \downarrow }$. Consider the
fact that the occupation number in the intra-dot level $\alpha \sigma $ is
mainly effected by the lead (i.e. electron reservoir) $\alpha \sigma $. The
grand thermodynamic potential $\Omega $ at the zero temperature is
\begin{eqnarray}
\Omega (\vec{N}) &=&E_{T}(\vec{N})-N_{L\uparrow }\mu _{L\uparrow
}-N_{L\downarrow }\mu _{L\downarrow }  \notag \\
&&-N_{R\uparrow }\mu _{R\uparrow }-N_{R\downarrow }\mu _{R\downarrow }.
\end{eqnarray}
In the present system, the electron occupation number can change
with the levels $\epsilon_L$ and $\epsilon_R$. This is a grand
canonical ensemble. Then the stablest state is one whose grand
thermodynamic potential $\Omega $ has the minimal values, and can be
found straightforwardly. For the sake of convenience and intuition,
we introduce the electrochemical potentials $\mu _{QD\alpha \sigma
}$, following Ref.\cite{ref10}. $\mu _{QD\alpha \sigma }$ of the
level $\alpha \sigma $ is well defined, for example,
\begin{equation}
\mu _{QDL\uparrow }(\vec{N})=E_{T}(\vec{N})-E_{T}(N_{L\uparrow
}-1,N_{L\downarrow },N_{R\uparrow },N_{R\downarrow }).
\end{equation}
Then the stablest states are the maximal values of $\vec{N}$ for which four $%
\mu _{QD\alpha \sigma }(\vec{N})$ are less than the corresponding chemical
potentials $\mu _{\alpha \sigma }$. If two states of $\vec{N}$, e.g. $\vec{N}%
=(0,0,1,0)$ and $(0,0,0,1),$ satisfy the above four equations, they are
assumed to have the same probability to exist. A detailed analysis of $\mu
_{QD\alpha \sigma }$ versus the parameters $\epsilon _{L}$ and $\epsilon
_{R} $ leads to establish the same charge stability diagram as shown in
Fig.3a. In fact, the electrochemical potentials $\mu _{QD\alpha \sigma }$
are equal to the equivalent levels in the preceding paragraph. For example,
\begin{eqnarray*}
\mu _{QDL\uparrow }(1,0,1,0) &=&\mu _{QDL\uparrow }(1,0,0,1)=\mu
_{QDL\downarrow }(0,1,1,0) \\
&=&\mu _{QDL\downarrow }(0,1,0,1)=\epsilon _{L}+U_{ex}=\tilde{\epsilon}_{L}.
\end{eqnarray*}
In particular, there are only four equivalent levels, which are less than
the numbers of $\mu _{QD\alpha \sigma }$. So it is convenient and intuitive
to use the equivalent levels to deduce the stability diagram.

With the spin-polarized stability diagram in mind, we turn to calculate the
(charge) current $J$ induced by the spin bias. Fig.2d shows the current $J$
as a function of the levels $\epsilon _{L}$ and $\epsilon _{R}$. The current
becomes quite large when both the left and right dots are spin polarized in
the case of $-V<\tilde{\epsilon}_{L}=\epsilon _{R}+U_{in}<V$. The physical
origin of generation of the current has been explained in detail in the
introduction and as shown in Fig.1. We can establish a relation between the
charge current and the spin bias in the two leads. In this way, we can
detect the spin bias $V$ by measuring the current $J$. In the following we
calculate the current for various parameters. Fig.4a shows the current $J$
versus the spin bias $V$ for the inter-dot interaction $U_{ex}=5$. While $V=0
$, $J$ is zero exactly. With the increase of $V$ from zero, the current $J$
first increases, reaches at a maximum, and then drops. $J$ keeps a
relatively large value even if $V$ is comparable with the e-e interaction
energy $U_{in} $. The origin of the drop is that the spin-polarizations in
two dots decay while the current flows through the DQDs at the large $V$. In
the absence of the inter-dot e-e interaction $U_{ex}$, i.e. $U_{ex}=0$, the
current increases monotonously with the spin bias $V$ (see Fig.4b). In this
case the current $J$ and the spin bias $V$ have a one-to-one correspondence.
Therefore the spin bias $V$ can be deduced straightforward from the measured
current. Fig.4c shows the current $J$ as a function of the right-dot's level
$\epsilon _{R}$. When $\epsilon _{R}+U_{in}$ departs $\tilde{\epsilon}_{L}$
over a few $\Gamma _{\alpha }$ (e.g. $|\epsilon _{R}+U_{in}-\tilde{\epsilon}%
_{L}|>3\Gamma _{\alpha }$), $J$ becomes very small because the tunneling
process in Fig.1a suppresses quickly when $\epsilon _{R}+U_{in}$ is not in
alignment with $\tilde{\epsilon}_{L}$ . On the other hand, the tunneling
process in Fig.1a occurs frequently and $J$ becomes large when $\epsilon
_{R}+U_{in}$ is located near $\tilde{\epsilon}_{L}$. However, when $\epsilon
_{R}+U_{in}=\tilde{\epsilon}_{L}$, $J$ may drop slightly and a dip emerges
in the curve of $J$-$\epsilon _{R}$, because the spin-polarization $\Delta
n_{\alpha }$ is suppressed at the point. Fig.4d displays the current $J$ as
a function of temperature $T$. Here $J$ depends on the temperature $T$
slightly, and is quite large when $T<V$.

\section{Charge bias in an open circuit}

In the preceding section, we calculated the charge current through a DQD
induced by a pure spin bias. In an open circuit, the situation will be
changed. At the time that a spin bias is turned on, a charge current will
circulate. For an open circuit, the extra charge will accumulate in the two
leads until the system reaches at a balance. As a result an extra charge
bias $V_{e}$ instead of a charge current will be generated while the charge
current vanishes. In this case, combination of the the spin bias $V$ and the
induced charge bias $V_{e}$ will give the spin-dependent chemical potentials
$\mu _{\alpha \sigma }$ in the two leads
\begin{subequations}
\begin{eqnarray}
\mu _{L\uparrow } &=&+V+V_{e}, \\
\mu _{L\downarrow } &=&-V+V_{e}, \\
\mu _{R\uparrow } &=&-V-V_{e}, \\
\mu _{R\downarrow } &=&+V-V_{e}.
\end{eqnarray}%
The bias $V_{e}$ can be determined by the condition of
\end{subequations}
\begin{equation}
J=0
\end{equation}%
in equilibrium for an open circuit. Figs. 5a and 5b gives the bias $V_{e}$
and $V_{e}/V$ versus the spin bias $V$ in the presence and absence of the
intra-dot Coulomb interaction $U_{ex}$. $|V_{e}|$ and $|V_{e}/V|$ increase
monotonously with $V$ regardless of the value of $U_{ex}$. This is different
from the curve of $J$-$V$, in which $J$ drops down for a large $V$ while $%
U_{ex}\not=0$ (see Fig.4a). This illustrates that it is more efficient to
measure the induced bias $V_{e}$ than to measure the induced current $J$.
Fig.5c shows the bias $V_{e}$ as a function of the level $\epsilon _{R}$.
The bias $|V_{e}|$ always has a large value (e.g. $|V_{e}/V|>0.1$), even if $%
\epsilon _{R}+U_{in}$ is far away from $\tilde{\epsilon}_{L}$. Notice that
the current $J$ is relatively small when $|\epsilon _{R}+U_{in}-\tilde{%
\epsilon}_{L}|>3\Gamma _{\alpha }$ (see Fig.4c). The transmission
coefficient (or the conductance) is also very small in this region.
Correspondingly, $V_{e}$ in an open circuit is still large.
Therefore the induced bias $V_{e}$ can be measured in an more
extensive region. Fig.5d gives the temperature $T$ dependence of the
bias $V_{e}$, which is almost independent of the temperature $T$.
Finally, we emphasize that $|V_{e}/V|$ is usually larger than $0.1$
regardless of the values of the parameters $V$, $\epsilon _{L}$,
$\epsilon _{R}$, $T$, etc. In the current technology, the bias in
the order of $0.1nV$ is measurable in experiment.\cite{add21}
Therefore, if the spin bias $V$, i.e. the difference of the spin-up
and spin-down chemical potentials $(\mu _{L\uparrow }-\mu
_{L\downarrow })/e$, reaches to $1nV$, the induced bias in the
present calculation is large enough to be measured in experiment.

\section{conclusions}

In summary, we investigated the electron transport driven by a spin
bias or pure spin current through a non-magnetic DQD. Except for the
spin-unpolarized domains, several spin-polarized domains are found
in the stability diagram with respect to the energy levels of two
quantum dots. When both of the left and right dots are spin
polarization, a large charge current $J$ can be induced by applying
of the pure spin bias. In particular, in an open circuit, the charge
bias is induced to balance the spin bias, and is measurable in an
extensive range of the parameters. Physically, a pure spin bias may
drive electrons with different spin in opposite direction. If the
system possesses the left-right symmetry or parity and does not
break the time reversal symmetry it will circulate a pure spin
current (or spin accumulation in an open circuit). When the energy
levels in the two dots are not equal, the left-right symmetry or
parity of the system is broken. A spin bias and strong Coulomb
interaction can produce two spin polarized states in the two dots as
we discussed in the spin-polarized charge stability diagram. As a
result the currents with different spins in opposite directions will
not be equal any more. Consequently this pure spin bias generates a
charge current through the DQD. This property may provide a
practical approach to detect the spin bias in DQD by measuring the
charge bias or charge current.

\section*{Acknowledgments}

We acknowledge the financial support from NSF-China under Grant
Nos. 10525418, 10734110, and 60776060 (Q.F.S.), and the Research
Grant Council of Hong Kong under Grant No.: HKU 7041/07P (S.Q.S.).
Q.F.S. would like to thank Dr. W. Long for many helpful
discussions.

\begin{figure}[tbph]
\caption{ (color online) (a) ((b)) The schematic plots illustrate an spin-up
(spin-down) electron tunneling from the left (right) to the right (left)
lead. }
\label{fig::fig1}
\end{figure}

\begin{figure}[tbph]
\caption{(color online) The spin polarization $\Delta n_{L}$ at the left dot
(a), $\Delta n_{R}$ at the right dot (b), $n_{L}+n_{R}/2$ (c), and the
current $J$ (d) as a function of the energy levels $\protect\epsilon _{L}$
and $\protect\epsilon _{R}$ in the two quantum dots. The parameters are: $%
\Gamma _{L}=\Gamma _{R}=0.3$, $t_{c}=T=0.1$, $U_{in}=20$, $U_{ex}=5$, and $%
V=1$. }
\label{fig::fig2}
\end{figure}

\begin{figure}[tbph]
\caption{ (color online) (a) Schematic stability diagram of the DQD under
the finite spin bias $V$. The thin dashed lines are the stability diagram
while $V=0$. (b), (c), and (d) are location schematics of energy levels in
the $(0,1)$ , $(\uparrow ,1)$, and $(0,\downarrow )$ domains, respectively. }
\label{fig::fig3}
\end{figure}

\begin{figure}[tbph]
\caption{The current $J$ vs. the spin bias $V$ for two inter-dot
interactions $U_{ex}=5$ (a) and $U_{ex}=0$ (b). (c) The current $J$ vs. the
level $\protect\epsilon _{R}$ and (d) the current $J$ vs. the temperature $T$%
. The solid, dashed, and dotted curves are for levels at $\tilde{\protect%
\epsilon}_{L}=0$ and $\protect\epsilon _{R}=-20$, $\tilde{\protect\epsilon}%
_{L}=0.5$ and $\protect\epsilon _{R}=-19.5$, and $\tilde{\protect\epsilon}%
_{L}=1$ and $\protect\epsilon _{R}=-19$, respectively. The other parameters
are the same as in Fig.2. }
\label{fig::fig4}
\end{figure}

\begin{figure}[tbph]
\caption{The induced charge bias $V_{e}$ (thin curves) and $V_{e}/V$ (thick
curves) vs. the spin bias $V$ for inter-dot interaction $U_{ex}=5$ (a) and $%
U_{ex}=0$ (b). (c) $V_{e}$ (i.e. $V_{e}/V$) vs. the level $\protect\epsilon %
_{R}$ and (d) $V_{e}$ (i.e. $V_{e}/V$) vs. the temperature $T$. The solid,
dashed, and dotted curves are for levels at $\tilde{\protect\epsilon}_{L}=0$
and $\protect\epsilon _{R}=-20$, $\tilde{\protect\epsilon}_{L}=0.5$ and $%
\protect\epsilon _{R}=-19.5$, and $\tilde{\protect\epsilon}_{L}=1$ and $%
\protect\epsilon _{R}=-19$, respectively. The other parameters are the same
as in Fig.2 }
\label{fig::fig5}
\end{figure}


\begin{references}
\bibitem{note21}
Here we took the approximation: $\langle \hat{n}_{L\sigma}
\hat{n}_{R\sigma'} \rangle = \langle \hat{n}_{L\sigma} \rangle
\langle \hat{n}_{R\sigma'} \rangle = n_{L\sigma}n_{R\sigma'}$, and
$\langle \hat{n}_{\alpha\uparrow} \hat{n}_{\alpha\downarrow}
\rangle =0$ while $0<n_{\alpha}<1$ and $\langle
\hat{n}_{\alpha\uparrow} \hat{n}_{\alpha\downarrow} \rangle
=n_{\alpha} -1$ while $1<n_{\alpha}<2$, where
$\hat{n}_{\alpha\sigma} \equiv
d^{\dagger}_{\alpha\sigma}d_{\alpha\sigma}$. Because of
$V<U_{in}$, the fluctuation of the occupation number is
less than one in each dot, and the approximation should be
reasonable.
\bibitem{note}
The occupation numbers of electron, in particular $n_L +n_R/2$, can be
measured by using the charge sensing techniques, i.e.,  using a
quantum point contact (QPC) nearby the DQDs device to measure the
conductance of the QPC. For the detail, see Refs.\cite{ref10,ref11}.
\end{references}
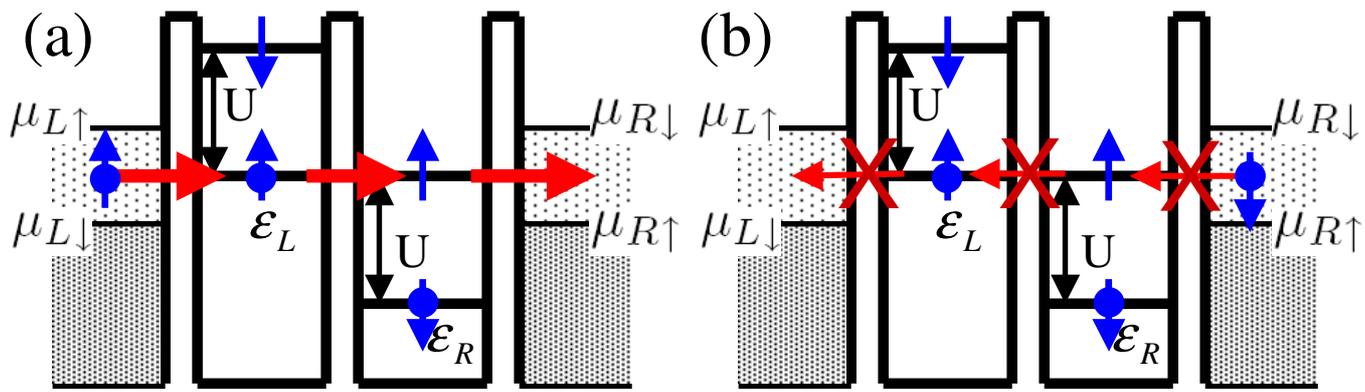
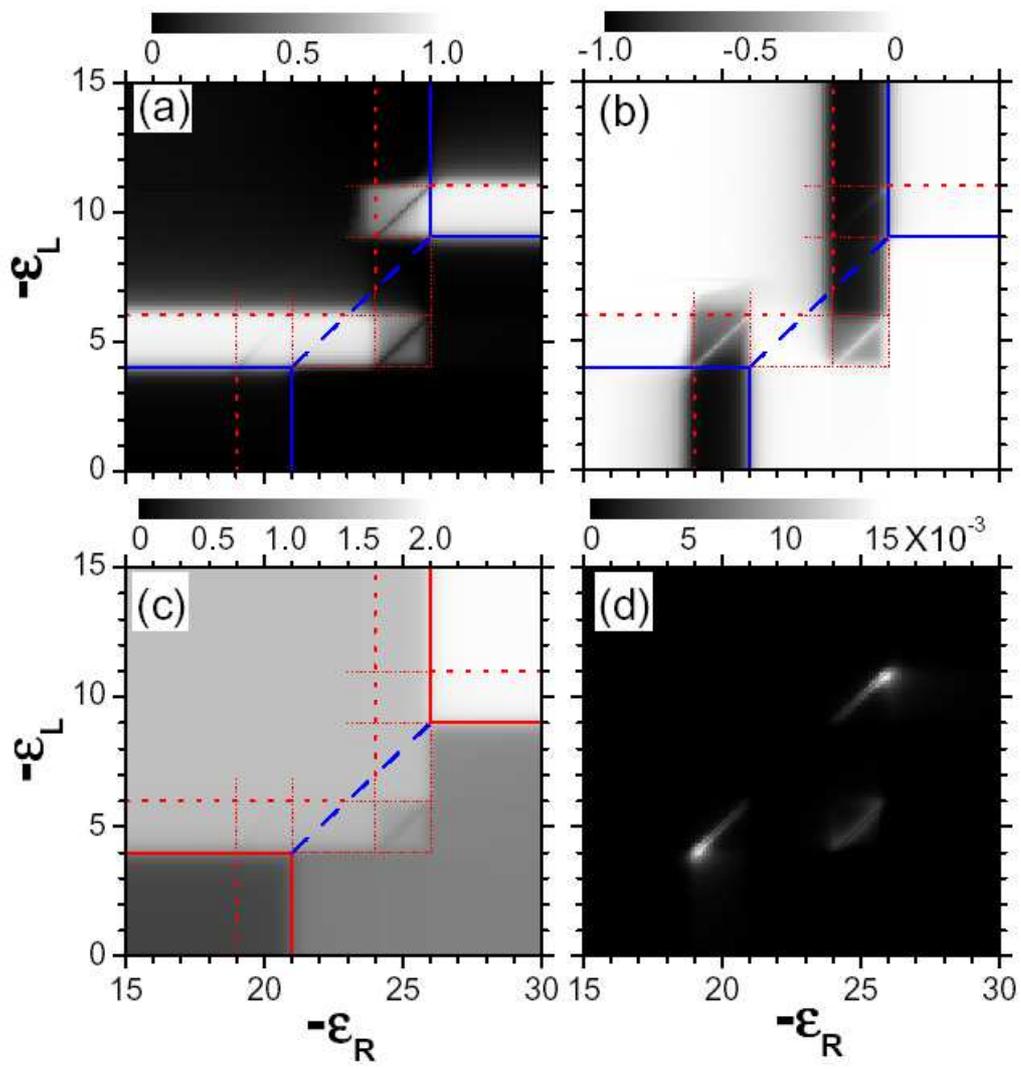
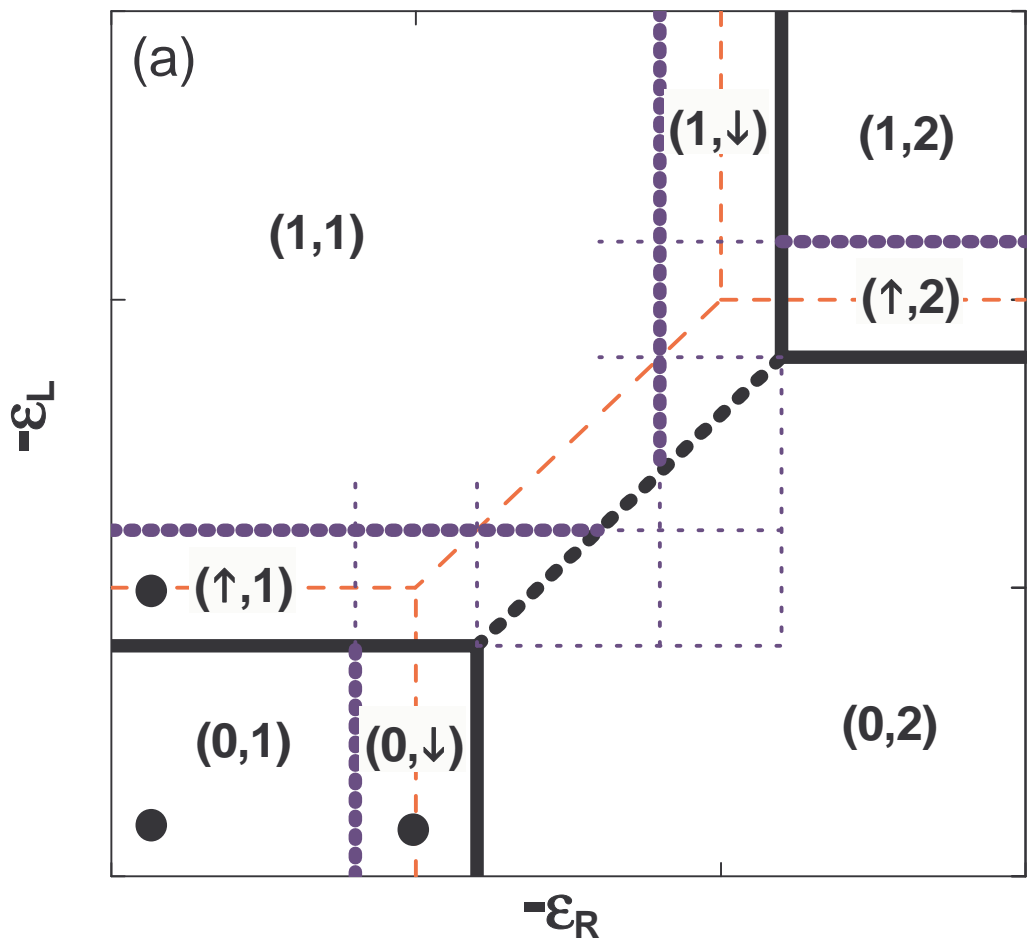
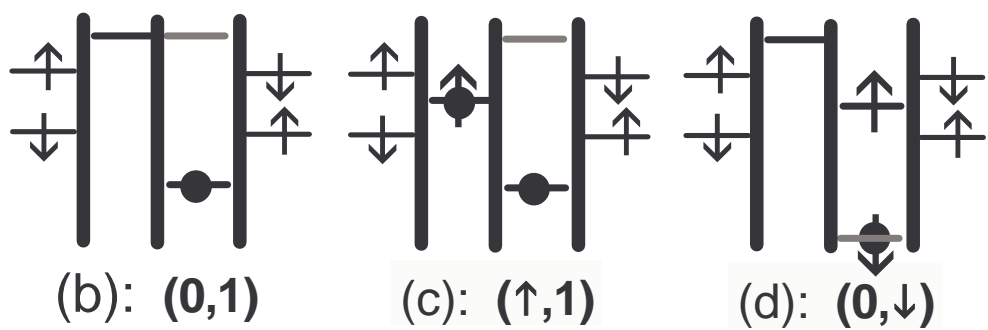
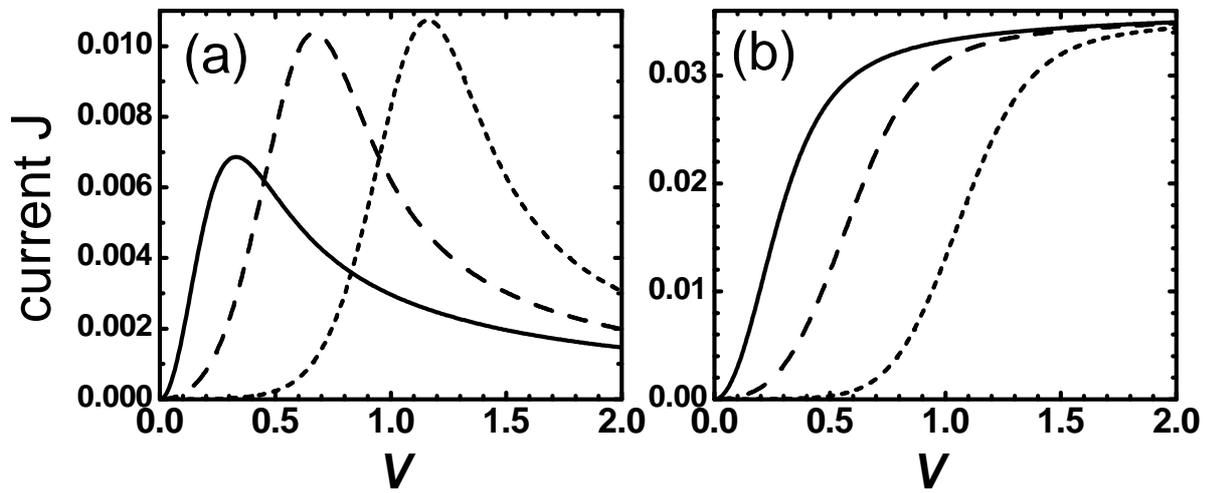
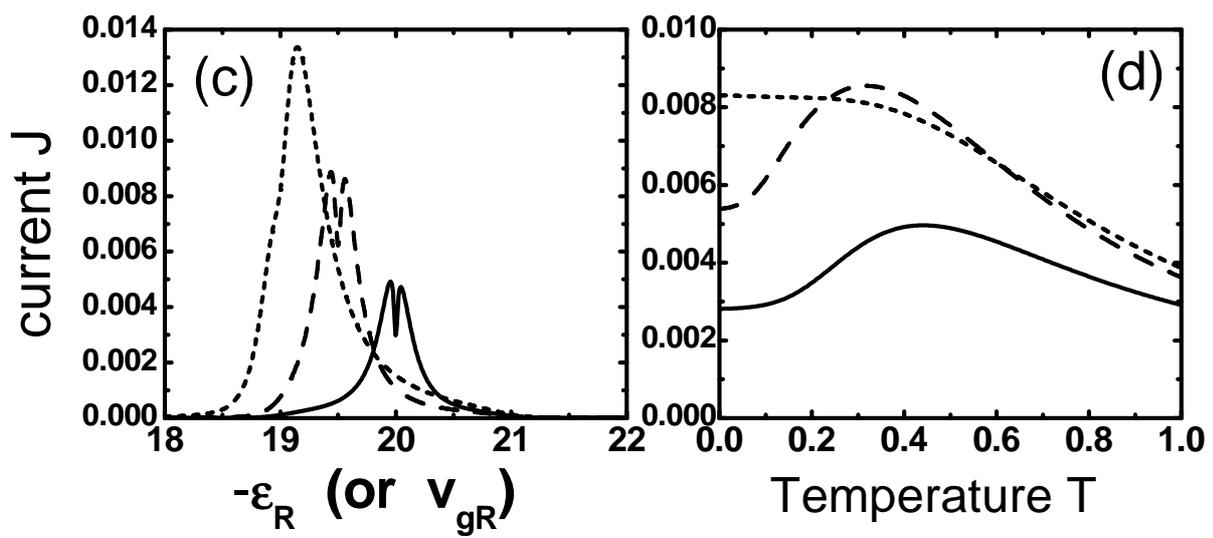
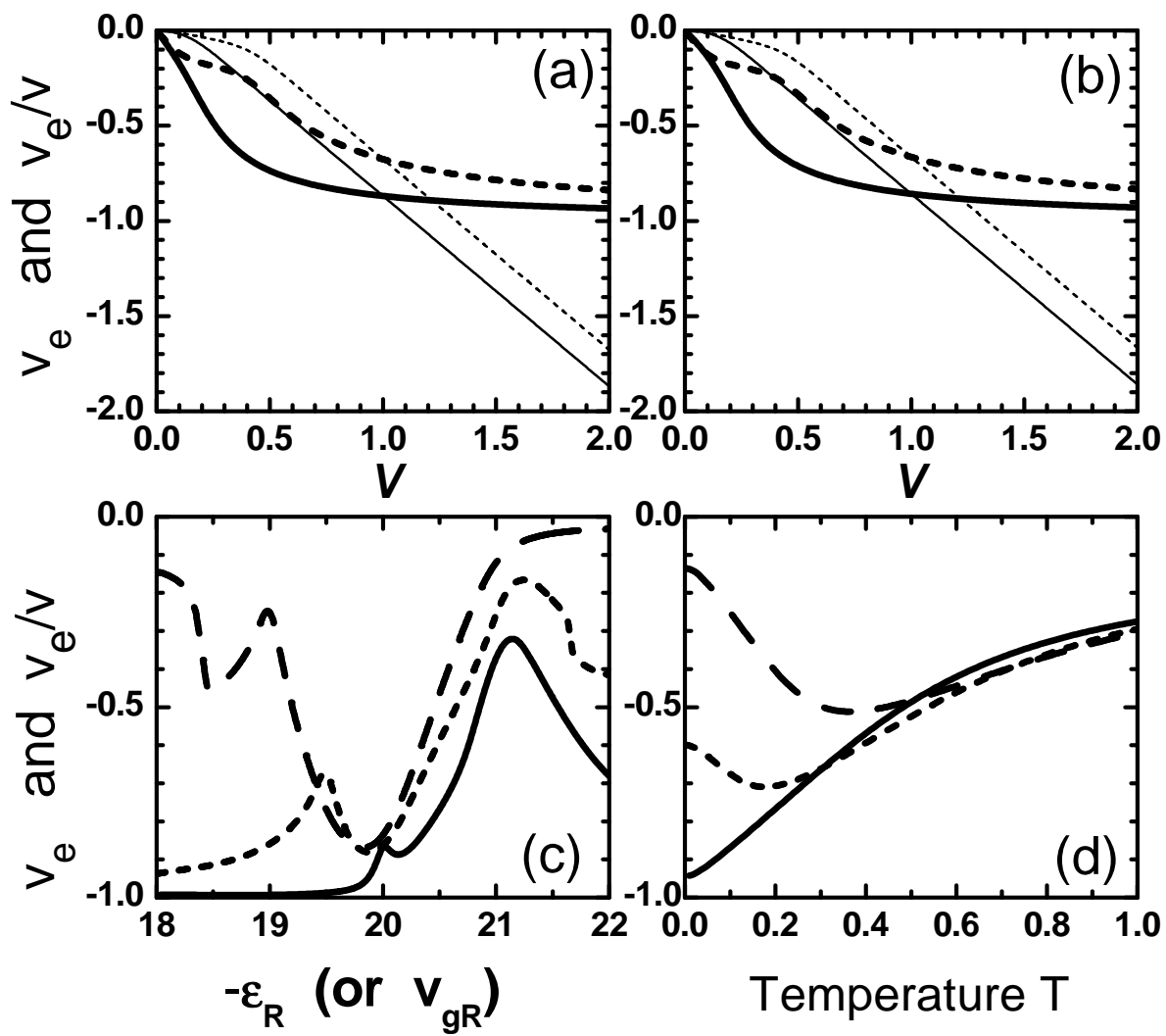
\end{document}